\documentclass[iop,apjl,numberedappendix]{emulateapj}
\usepackage[latin1]{inputenc}
\usepackage{amssymb}
\usepackage{amsmath}
\usepackage{graphicx}
\usepackage{xspace}
\usepackage{natbib}
\usepackage{url}
\usepackage{paralist}
\usepackage{epsfig}

\def\cm{{\rm\thinspace cm}}

\def\erg{{\rm\thinspace erg}}
\def\eV{{\rm\thinspace eV}}

\def\K{{\rm\thinspace K}}
\def\keV{{\rm\thinspace keV}}
\def\km{{\rm\thinspace km}}

\def\Msun{\hbox{$\rm\thinspace M_{\odot}$}}

\def\s{{\rm\thinspace s}}

\def\yr{{\rm\thinspace yr}}

\def\ergcmps{\hbox{$\erg\cm\s^{-1}\,$}}

\def\ergps{\hbox{$\erg\s^{-1}\,$}}

\def\kmps{\hbox{$\km\s^{-1}\,$}}

\def\Msunpyr{\hbox{$\Msun\yr^{-1}\,$}}

\def\pcmsq{\hbox{$\cm^{-2}\,$}}

\def\h18{\hbox{H1821$+$643\,}}
\def\chandra{\hbox{{\it Chandra}}}
\def\suzaku{\hbox{{\it Suzaku}}}

\slugcomment{ }

\shorttitle{X-ray spectrum of H1821$+$643}
\shortauthors{C.~S.~Reynolds et al.}

\begin{document}

\title{The X-ray spectrum of the cooling-flow quasar H1821$+$643 : \\ A massive black hole feeding off the intracluster medium}

\author{Christopher~S.~Reynolds\altaffilmark{1,2}, Anne~M.~Lohfink\altaffilmark{1,3}, Arif~Babul\altaffilmark{4}, Andrew~C.~Fabian\altaffilmark{3}, Julie~Hlavacek-Larrondo\altaffilmark{5,6,7}, Helen~R.~Russell\altaffilmark{3} and Stephen~A.~Walker\altaffilmark{3}}
\email{chris@astro.umd.edu}
\altaffiltext{1}{Department of Astronomy, University of Maryland, College Park, MD 20742-2421}
\altaffiltext{2}{Joint Space-Science Institute (JSI), College Park, MD 20742-2421}
\altaffiltext{3}{Institute of Astronomy, Madingley Road, Cambridge, CB3 OHA, UK}
\altaffiltext{4}{Dept. of Physics and Astronomy, University of Victoria, Victoria, British Columbia V8P~5C2, Canada}
\altaffiltext{5}{D\'epartement de Physique, Universit\'e de Montr\'eal, C.P. 6128, Succ. Centre-Ville, Montr\'eal, Qu\'ebec H3C 3J7, Canada}
\altaffiltext{6}{Kavli Institute for Particle Astrophysics and Cosmology, Stanford University, 382 Via Pueblo Mall, Stanford, CA 94305-4060}
\altaffiltext{7}{Department of Physics, Stanford University, 452 Lomita Mall, Stanford, CA 94305-4085}
\begin{abstract}
\noindent We present a deep \suzaku\ observation of \h18, an extremely rare example of a powerful quasar hosted by the central massive galaxy of a rich cooling-core cluster of galaxies.  Informed by previous \chandra\ studies of the cluster, we achieve a spectral separation of emission from the active galactic nucleus (AGN) and the intracluster medium (ICM).   With a high degree of confidence, we identify the signatures of X-ray reflection/reprocessing by cold and slowly moving material in the AGN's immediate environment.   The iron abundance of this matter is found to be significantly sub-solar ($Z\approx 0.4Z_\odot$), an unusual finding for powerful AGN but in line with the idea that this quasar is feeding from the ICM via a Compton-induced cooling flow.  We also find a subtle soft excess that can be described phenomenologically (with an additional black body component) or as ionized X-ray reflection from the inner regions of a high inclination ($i\approx 57^\circ$) accretion disk around a spinning ($a>0.4$) black hole.  We describe how the ionization state of the accretion disk can be used to constrain the Eddington fraction of the source.  Applying these arguments to our spectrum implies an Eddington fraction of 0.25--0.5, with an associated black hole mass of $3-6\times 10^9\Msun$.  
\end{abstract}



\section{Introduction}\label{intro}

\noindent Although rare, the brightest cluster galaxies (BCGs) at the centers of cooling-core clusters of galaxies provide an important window into the physics of galaxy formation.  Most of the baryons in galaxy clusters reside in the hot intracluster medium (ICM) that produces an approximately hydrostatic atmosphere in the gravitational potential of the cluster.  The core of the ICM in 50\% of galaxy clusters have radiative cooling times below 1\,Gyr \citep{hudson:10a}.   Unchecked, radiative loses would induce flows of cooled gas into the BCG resulting in $100-1000\Msunpyr$ of star formation, far exceeding observed levels \citep{fabian:94a}.  It is now widely believed that mechanical heating of the ICM by a radio-loud active galactic nucleus (AGN) residing in the BCG is responsible for offsetting the ICM cooling, truncating the further growth of the BCG \citep{peterson:06a,fabian:12a}.  This conclusion is motivated by the fact that a very large fraction of BCGs in cool-core clusters host modest-power radio-loud AGN \citep{burns:90a}, and that imaging X-ray observations show clear signs of interaction between these AGN jets and the ICM \citep{fabian:00a,heinz:02a,birzan:04a}.

The most puzzling aspect of this AGN/ICM-feedback picture is the fueling of the AGN --- how does the AGN fueling rate (determined by conditions on parsec scales) lock into the right range required to offset radiative loses on the scale of the cluster core (10--100\,kpc)?   Are there systems where this regulatory feedback loop breaks down?    If so, what is their nature? 

Here, we study one such system where a stable AGN/ICM feedback may have broken down, at least temporarily and possible for an extended period.  \h18 ($z=0.297$) is an extremely rare example of a luminous broad-line quasar residing in the BCG of a rich cluster --- the only other comparable object in the local ($z<0.5$) Universe is the obscured quasar IRAS 09104$+$4109 \citep{osullivan:12a}.  Despite being classified as radio-quiet, \h18 is surrounded by a 300\,kpc FRI-type radio-structure suggesting either unusual jet activity in this optically-luminous quasar or a relic from a recent radio-galaxy phase.  \h18 shows no evidence for outflows in either \chandra\ grating studies \citep{fang:02a,mathur:03a} or FUSE UV spectroscopy \citep{oegerle:00a}.  

The ICM around \h18 was first clearly resolved by ROSAT \citep{hall:97a}, but detailed study had to wait for \chandra.   \cite{russell:10a} analyzed a moderately-deep \chandra\ exposure and, by carefully accounting for the (instrumentally) scattered emission from the quasar, showed that this is indeed a classical cool core cluster with temperature dropping from 7--8\,keV at $r=100$\,kpc down to 2--3\,keV within 20\,kpc.  More recently, using the same data, \cite{walker:14a} show that the \h18 cluster possesses an anomalously low core entropy when compared with other similar mass cool-core clusters.   They suggest that, if the quasar was even more luminous in the past, the entire ICM core within 80\,kpc could have been Compton-cooled by quasar radiation.  Indeed, it is possible that the current quasar ($L_{\rm bol}\sim 2\times 10^{47}\ergps$) is being fed by Bondi-accretion from a Compton-cooled ICM core.  Compton-cooling by the quasar radiation (with Compton-temperature $kT_C\sim 0.4\keV$) dominates over Bremsstrahlung cooling within the central 5\,kpc \citep{russell:10a}.    \cite{walker:14a} suggest that systems such as \h18 can get locked into a Compton-cooling fed state and hence grow ultramassive ($\gtrsim10^{10}\Msun$) black holes.  If correct, it represents a dramatic breakdown of the AGN/ICM-regulation picture thought to operate in most clusters.  

In this letter, we present an analysis of a deep observation of \h18 by the {\it Suzaku} X-ray Observatory.  Our focus is on the X-ray spectrum of the quasar itself, and builds on a body of previous work.   {\it Ginga} was the first to detected the iron-K complex in this source \citep{kii:91a}, {\it ASCA} showed the complex to be broadened \citep{yamashita:97a}, the {\it Chandra} High Energy Gratings detected the 6.4\,keV fluorescence line from the AGN \cite{fang:02a} and suggested the presence of a redshifted absorption line in the quasar spectrum \citep{yaqoob:05a}, and finally {\it XMM-Newton} isolated both the cold 6.4\,keV iron line and an ionized line that they tentatively associated with the ICM of the cluster \citep{Bailon:07a}.   

This Letter is organized as follows.  In \S\ref{sec:chandra} we discuss our analysis of the \chandra\ data from which we produce an ICM spectral model.  \S\ref{sec:suzaku} presents our analysis of the \suzaku\ observation, and we discuss the astrophysical implications of our findings in \S\ref{sec:discussion}.  Conclusions are drawn in \S\ref{sec:conclusions}.  Throughout this work, we assume standard {\it Planck} cosmology \citep{ade:13a}.  At a redshift of $z=0.297$, this places \h18 at a luminosity distance of 1.59\,Gpc, with a linear-angular conversion of 275\,kpc per arcmin.

\section{The Chandra view of the ICM}
\label{sec:chandra}

\begin{figure}[t]
\begin{center}
\psfig{figure=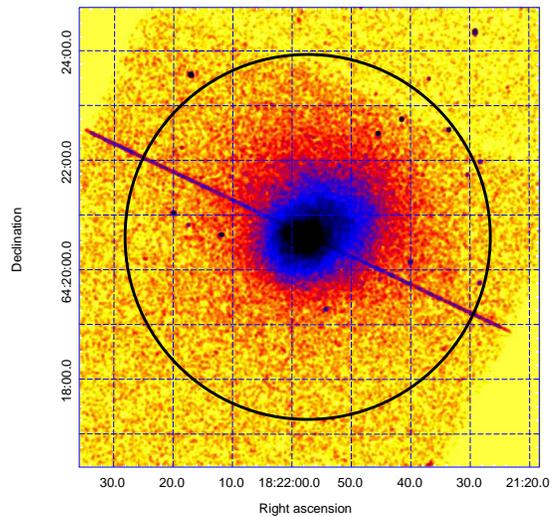,width=0.55\textwidth}
\caption{0.5--5\,keV \chandra\ ACIS-S3 image of \h18.  The image is dominated by extended ICM emission.  The prominent linear feature is the read-out streak from \h18 at the center of this cluster.  The black circle shows the extraction region that is used for our \suzaku\ spectral analysis.}
\label{fig:chandra}
\end{center}
\end{figure}

{\it Chandra} observed \h18 in imaging mode on four visits taken within the space of 9 days in April-2008 --- these data are described in detail by \cite{russell:10a} and \cite{walker:14a} who undertake a spatially-resolved study of the ICM.  Since we are interested in a global ICM spectrum over the (large) \suzaku\ extraction radius, we reanalyze the data thereby also employing the latest calibration and contaminant modeling.  

The data were obtained from the \chandra\ archives and reprocessed level-2 events files were created with the most recent software and calibrations (CIAOv4.6.1 and CALDBv4.5.9).  The merged 0.5--5\,keV \chandra\ image is shown in Fig.~\ref{fig:chandra}.  The image is dominated by the extended emission of the intracluster medium (ICM), although the read-out streak from the luminous central AGN is clearly visible.  To obtain a global spectrum of the ICM, we extract a spectrum from a circular region of radius 200\,arcsec (black circle on Fig.~\ref{fig:chandra}).  We exclude the innermost 5\,arcsec (radius) and the readout streak in order to isolate, as much as possible, the ICM spectrum from the AGN emission.  Background is extracted from an annulus centered on the cluster with inner and outer radius of 200\,arc sec and 240\,arc sec respectively (excluding the readout streak and a bright source close to RA 18h22m17s, Dec $+$64d23m32s).

The resulting ICM spectrum is modeled using the XSPECv12.8.1g package.  We find that a two-temperature thermal plasma ({\tt apec}) model modified by Galactic absorption (using {\tt tbabs} with $N_H=3.5\times 10^{20}\pcmsq$) provides an excellent description of the spectrum.  Even in this deep pointing, there are a small number of counts in the all-important iron-K band.   Hence, to respect the Poisson nature of the statistics we employ the Bayesian analysis ({\tt lstat}) capabilities of XSPEC.  The resulting fit gives the two temperatures as $kT_1=2.0_{-0.7}^{+2.5}\keV$ and $kT_2=9.3\pm 1.2\keV$ and the (common) ICM abundance as $Z=0.20\pm 0.05Z_\odot$.  The higher temperature component strongly dominates at all energies.  The total 0.5--10\,keV (rest frame) luminosity of the ICM is $2.55\times 10^{45}\ergps$. 

\section{A Suzaku study of the AGN spectrum}
\label{sec:suzaku}

{\it Suzaku} has observed \h18 on three occasions.  Here, we focus on the two observations from April-2013 that essentially form a single quasi-continuous stare with total XIS good exposure time of 375\,ks (ObsIDs 708037010, 708037020).  Data from all three operating XISs (XIS0, XIS1 and XIS3) were reprocessed and subjected to standard screening criteria using the {\tt aepipeline} tool within the HEASOFTv6.15.1 suite.  From the cleaned data, XIS spectra and lightcurves were extracted from a 200\,arcsec radius circular aperture centered on \h18.  Background spectra and lightcurves were extracted from source free regions around the edges of the XIS chips.  No time variability of the source was found in any of the detectors and hence our discussion will address exclusively the spectral domain.

To maximize our flexibility to handle calibration issues, we do not combine XIS spectra. We jointly fit six XIS spectra (XIS0/XIS1/XIS3 spectra from each of the two ObsIDs) in the spectral range 0.7--10\,keV (observed frame).  We also permit multiplicative offsets between different instruments to allow for flux-calibration errors.  

\begin{figure}[t]
\begin{center}
\psfig{figure=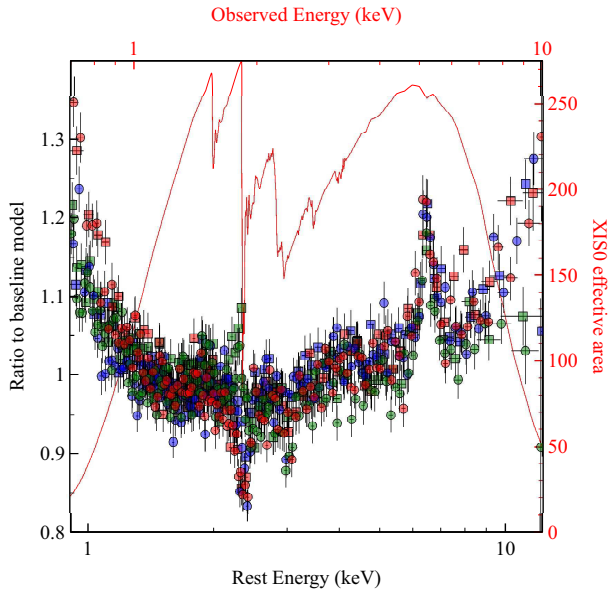,width=0.45\textwidth}
\caption{Ratio of XIS spectra to a baseline model consisting of ICM (fixed to \chandra\ derived parameters) and the best-fitting simple power-law.  Data are shown separately for the three XIS (red=XIS0, green=XIS1, blue=XIS3) and two ObsIDs (circles=Obs708037010, squares=708037020).  Comparison with the XIS0 effective area curve (red thin line) reveals that the sharp features bracketing 2\,keV (observed) are very likely calibration in nature.}
\label{fig:xis_porat}
\end{center}
\end{figure}

\begin{figure*}[t]
\begin{center}
\hbox{
\psfig{figure=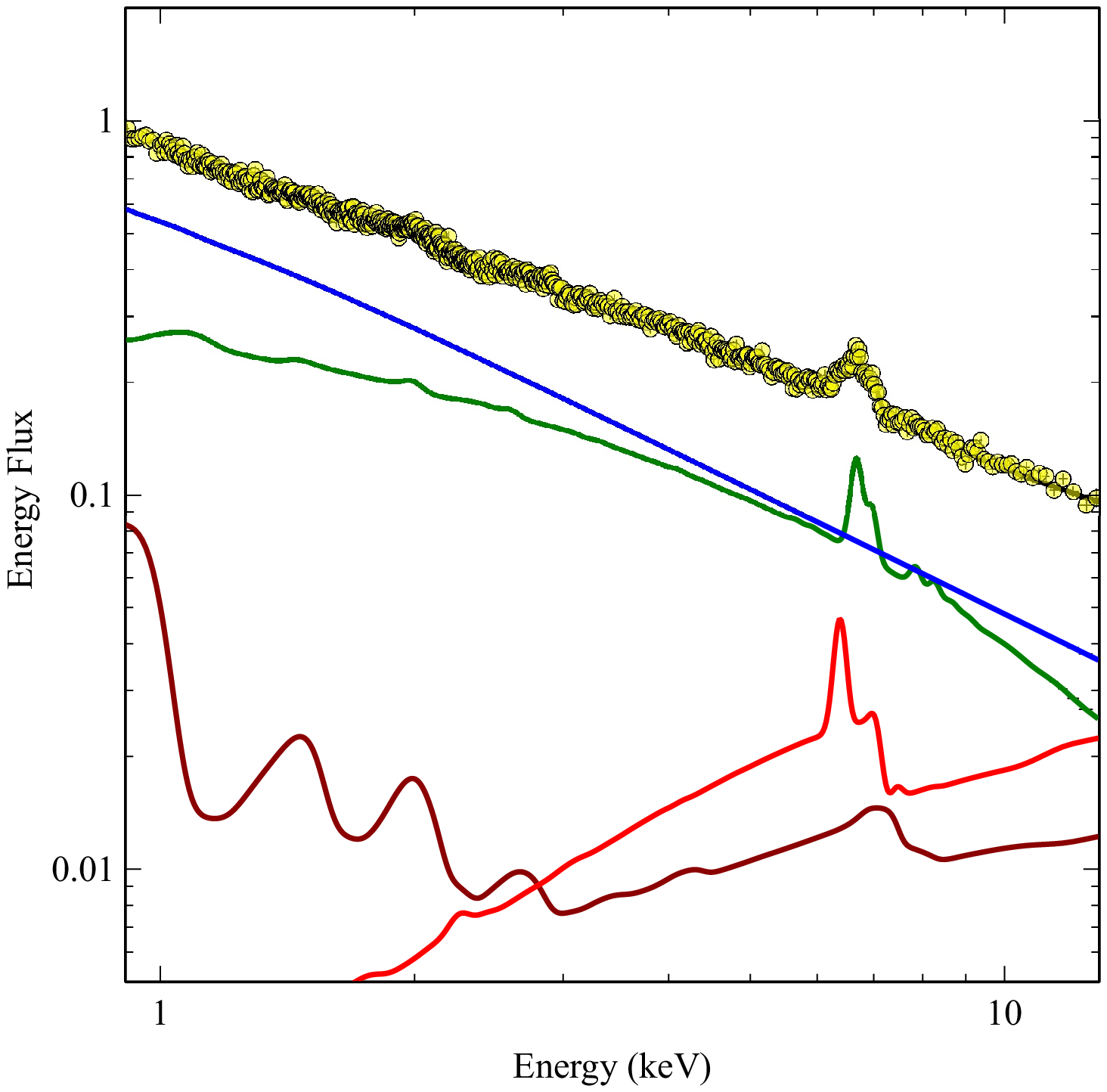,width=0.4\textwidth}
\hspace{1cm}
\psfig{figure=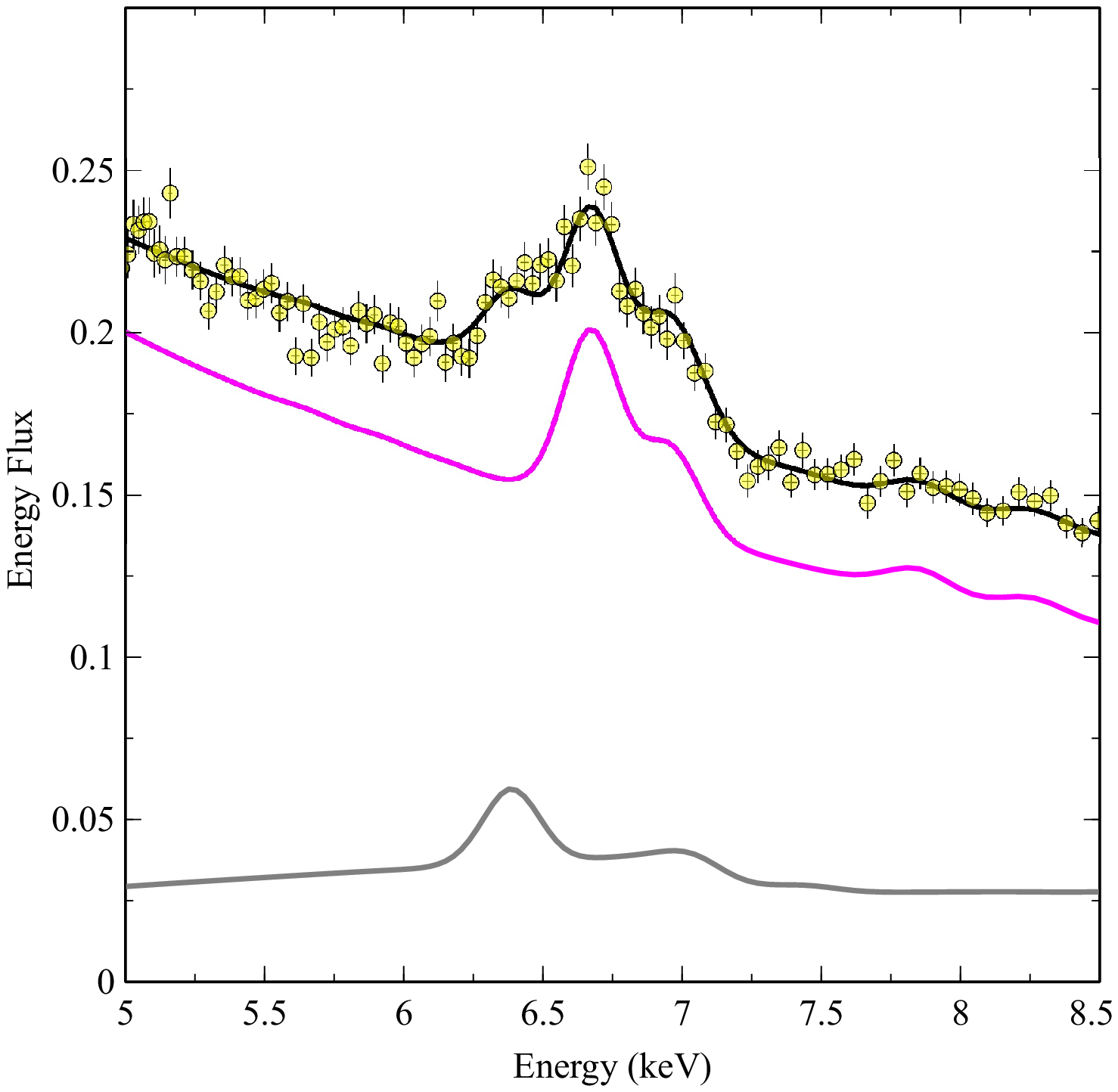,width=0.4\textwidth}
}
\caption{ {\it Left panel : }Effective-area corrected XIS0+XIS3 spectra together with our final spectral model (ICM+PL+REFL+DISK).  Coloring of model decomposition (from top to bottom); blue=PO, green=ICM, red=REFL, brown=DISK.  For plotting purposes (but not for fitting), the spectra from these two backside-illuminated XISs have been combined.  {\it Right panel : }Zoom-in on the iron line complex, breaking out the contribution from AGN reflection (grey=REFL+DISK) from the other components (magneta=ICM+PO).  }
\label{fig:xis_fit}
\end{center}
\end{figure*}

Initially, we fit the XIS spectra with a model consisting of the two-temperature ICM component derived from the \chandra\ analysis (\S\ref{sec:chandra}) and an additional power-law (PL) component to present the AGN primary continuum.  Figure~\ref{fig:xis_porat} shows the ratio of the data to this model (with best-fitting photon index $\Gamma=2.03$).  Several features of this fit are immediately apparent.  Firstly, there is a strong and unmodeled iron-K$\alpha$ emission line at 6.4\,keV.  Note that this is in addition to the Fe25-K$\alpha$ and Fe26-K$\alpha$ lines associated with the ICM that are accounted for in the spectral model and hence do not show up in this ratio plot.  Secondly, the spectrum has unmodeled curvature, with residuals indicating both a soft excess below 1\,keV and a hard tail above 8\,keV (energies quoted in rest-frame of \h18).  Thirdly, there are clear calibration issues --- there are residuals at the 10\% level at energies coincident with the Silicon (detector) edge at 1.8\,keV and the Gold (mirror) edge at 2.2\,keV, indicating gain shifts in the detectors.  Furthermore, the XIS0 is discrepant at the softest energies and XIS1 is discrepant at the highest energies, again possibly due to gain shifts.  

The 6.4\,keV emission line is a signature of reflection/reprocessing of the primary hard X-ray continuum by cold, optically-thick material close to the AGN.   Hence, to our spectral model, we add a neutral and slow-moving X-ray reflection (REFL) component described using the {\tt pexmon} model of \cite{nandra:07a}.  We also permit the instrumental gain parameters to be fit parameters in an attempt to ``self-calibrate'' the data.  Furthermore, acknowledging that the \chandra\ calibration may not be perfect, and the \suzaku\ point spread function mixes emission from different spatial regions of the cluster, we are conservative and permit the temperature, abundance and normalization of the dominant ($\sim 9$\,keV) ICM component to freely fit.  

\begin{table*}[t]
\begin{center}
\begin{tabular}{lcc}\hline\hline
Spectral Model & Parameters & $\chi^2/{\rm dof}$\\\hline
ICM+PL$^*$ & $\Gamma=2.03\pm 0.01$ & 11365/7795 (1.46) \\\hline

ICM+PL+REFL & $kT_2=8.4^{+0.5}_{-0.3}\keV, Z_{\rm Fe,ICM}=0.27\pm 0.03Z_\odot$ & 8302/7778 (1.067)\\ 
 & $\Gamma=2.08_{-0.05}^{+0.02}, {\cal R}_{\rm cold}=1.37^{+0.27}_{-0.23}, Z_{\rm Fe,REFL}=0.39^{+0.03}_{-0.06}$ & \\\hline
 
 ICM+PL & $kT_{\rm ICM}=8.75^{+0.20}_{-0.25}, Z_{\rm Fe,ICM}=0.36^{-0.03}_{+0.05}$  & 8287/7775 (1.066)  \\
+LINE+HUMP & $E_{\rm line}=6.41\pm 0.02\keV, \sigma_{\rm line}=60^{+35}_{-14}\eV, W_{\rm line}=46^{+10}_{-7}\eV$ & \\
 & $\Gamma=2.07\pm 0.02, {\cal R}=0.77^{+0.07}_{-0.11}, Z_{\rm Fe,REFL}=0.12^{+0.05}_{-0.02}$  & \\\hline
 
 ICM+PL & $kT_{\rm ICM}=8.0^{+0.4}_{-0.5}\keV, Z_{\rm Fe,ICM}=0.33^{+0.06}_{-0.04}Z_\odot$  & 8175/7776 (1.051)\\
 +REFL+BBOD& $\Gamma=2.01\pm 0.03, {\cal R}=1.0\pm 0.2, Z_{\rm Fe,REFL}=0.45^{+0.11}_{-0.08}Z_\odot$  & \\
 & $kT_{\rm bb}=79^{+13}_{-12}\eV$  & \\\hline
 
ICM+PL & $kT_{\rm ICM}=7.8\pm 0.4\keV, Z_{\rm Fe,ICM}=0.20\pm 0.02Z_\odot$ & 8180/7773 (1.052)\\
 +REFL+DISK & $\Gamma=2.13\pm 0.03, {\cal R}=1.46\pm0.02, Z_{\rm Fe,REFL}=0.41^{+0.11}_{-0.06}Z_\odot$ & \\
 & $i=57\pm 3,h<11,r_{\rm in}<3.6, \log\xi=1.68\pm 0.2,{\cal R}_{\rm disk}=0.79^{+0.35}_{-0.12}$ & \\\hline
 
 ICM+PL+REFL & $kT_{\rm ICM}=7.6^{+0.5}_{-0.7} \keV, Z_{\rm Fe,ICM}=0.21^{+0.03}_{-0.02}Z_\odot$ & 8142/7771 (1.048)\\
 +DISK+BBOD & $\Gamma=2.03^{+0.02}_{-0.01}, {\cal R}=1.15^{+0.18}_{-0.35}, Z_{\rm Fe,REFL}=0.46^{+0.20}_{-0.08}Z_\odot$ & \\
  & $i=51^{+4}_{-5},h<17,r_{\rm in}<7.0, \log\xi=2.00^{+0.05}_{-0.24},{\cal R}_{\rm disk}=0.41^{+0.16}_{-0.15}$  & \\
   & $kT_{\rm bb}=0.13\pm 0.02\keV$  & \\\hline\hline
 \end{tabular}
\end{center}
\caption{Parameters for \suzaku\ spectral fits.   For the first model (flagged by *), ICM parameters are fixed at those derived from \chandra\ and the gain parameters of the XISs were held fixed at default values.  For the subsequent fits, the parameters (temperature, abundance and normalization) of the hot ICM component were allowed to fit freely, as were the six sets of XIS gain parameters (three XISs, two ObsIDs).  Errors are quoted at the 90\% confidence level for one interesting parameter.  }
\label{tab:fits}
\end{table*}%

This model (ICM+PL+REFL) is a much better description of the data (Table~\ref{tab:fits}).  The cold X-ray reflection accounts for the 6.4\,keV iron line and much of the hard tail, and small but significant ($<2\%$) gain shifts largely remove the calibration residuals at the silicon and gold edges and bring the different XIS detectors into agreement with each other across the energy band.  While the amplitude of the reflection is similar to that seen in many AGN (with a reflection fraction of ${\cal R}=1.4$), the fit demands sub-solar iron abundance for the X-ray reprocessing circumnuclear matter ($Z_{\rm Fe}\approx 0.4Z_\odot$; this is driven by the relative strength of the iron line to the hard tail).  Disaggregating the reflection component into an iron line (modeled with a Gaussian) and the Compton reflection continuum \citep[modeled using the {\tt pexrav} code, ][]{magdziarz:95a}, we find a line energy consistent with cold iron-K$\alpha$ fluorescence (6.40\,keV) that is marginally resolved ($\sigma\approx 60_{-14}^{+35}\eV$ corresponding to a full-width half-maximum [FWMH] velocity of $6600^{+3800}_{-2500}\kmps$).  This disaggregated model  (ICM+PL+LINE+HUMP) permits unphysical degrees of freedom, explaining the slight discrepancy in the values of ${\cal R}$ and $Z_{\rm Fe}$ as compared with the ICM+PL+REFL fit (Table~\ref{tab:fits}). Still, the need for sub-solar abundances is clear from both the form of the Compton hump and the low equivalent width of the iron line.

This fit still leaves a subtle unmodeled soft-excess below 1\,keV.  This cannot be due to an additional low-temperature ICM component as the required normalization of such a component would strongly conflict with our \chandra\ spectrum as well as requiring unphysical emission measures.   Phenomenologically, the soft excess can be described by a black-body (BBOD in Table~\ref{tab:fits}) with rest-frame temperature $kT_{\rm bb}\approx 80\eV$ ($T_{\rm bb}=9.3\times 10^5\K$).  However, the physical origin of such a component (which would dominate the unabsorbed 0.1--10\,keV luminosity) is unclear given that, even assuming a maximally spinning black hole, the innermost regions of the accretion disk will have temperature $T_{\rm bb}\lesssim 3\times 10^5\K$ \citep{shakura:73a}.  We note that our conclusions regarding the low-abundance of the X-ray reflector is unaffected by the inclusion of this phenomenological soft-excess.

Instead, we suggest that the soft excess arises by reflection from the inner/relativistic regions of an ionized accretion disk.  We add this component to the spectral model, employing the {\tt relxill\_lp} code \citep{garcia:14a} which describes the relativistically blurred reflection spectrum from a geometrically-thin disk around a Kerr black hole illuminated by an X-ray ``lamp-post'' on the black hole spin axis at height $h$.  Following \cite{reynolds:12a}, we assume that the inner accretion disk shares the same metallicity as the other circumnuclear reflecting gas.   Adding the disk reflection leads to a further highly significant improvement in the fit (Table~\ref{tab:fits}).  This model (ICM+PL+REFL+DISK) is shown against the (back-illuminated) XIS data in Fig.~\ref{fig:xis_fit} (left).   Figure~\ref{fig:xis_fit} (right) illustrates how the superior sensitivity and spectral resolution of the {\it Suzaku}/XIS permits the Fe25 and Fe26 ionized lines from the ICM, and the cold iron-K line from the AGN, to be distinguished.

We note that, even when the blackbody is used to describe the soft-excess, the fit is improved by the addition of the accretion disk ($\Delta\chi^2=33$ for 5 additional parameters; $F$-test null hypothesis probability of $<10^{-5}$). The constraints on the disk parameters in this composite soft-excess model are significantly poorer, however (last model in Table~\ref{tab:fits}).

\section{Discussion}
\label{sec:discussion}

Our \suzaku\ spectrum has permitted the cleanest isolation yet of X-ray reprocessing in the circumnuclear environment of this rare BCG-hosted quasar.  We find X-ray reflection from metal-poor ($Z\approx 0.4Z_\odot$), cold and slow-moving ($\lesssim 10^4\kmps$) material in the vicinity of the AGN.   We also find evidence for ionized X-ray reflection by the relativistic inner regions of the accretion disk.  We consider this detection of relativistic disk reflection to be tentative since it relies upon the modeling of subtle continuum curvature (i.e. soft and hard excesses) rather than, for example, an obvious broad iron line.  Still, taken at face value, our fiducial model suggests moderately ionized ($\xi\sim 50\ergcmps$) reflection from the innermost regions ($r_{\rm in}<3.6r_g$) of a high inclination ($i\approx 57^\circ$) accretion disk.  

\begin{figure*}
\centerline{
\psfig{figure=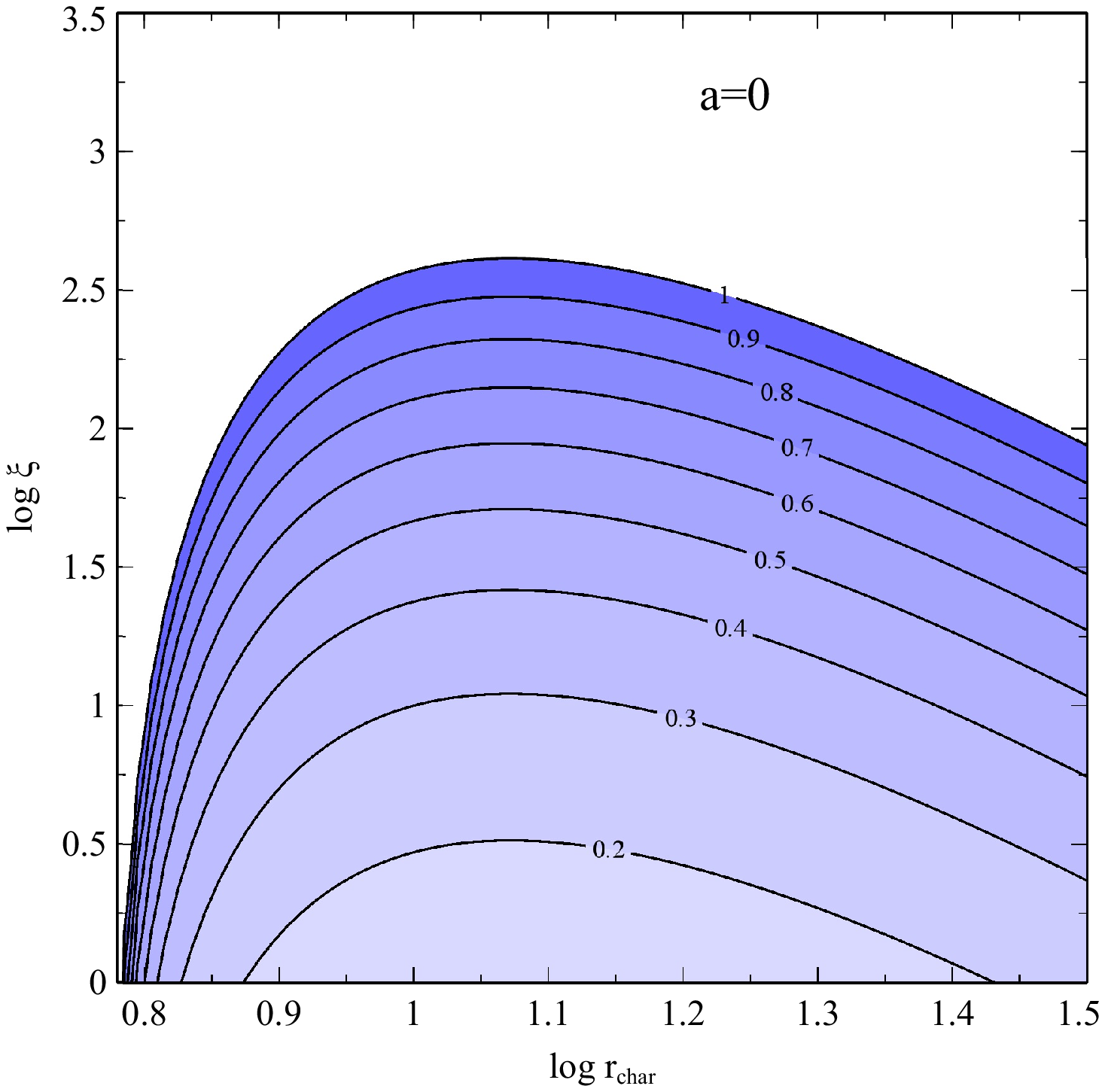,width=0.33\textwidth}
\psfig{figure=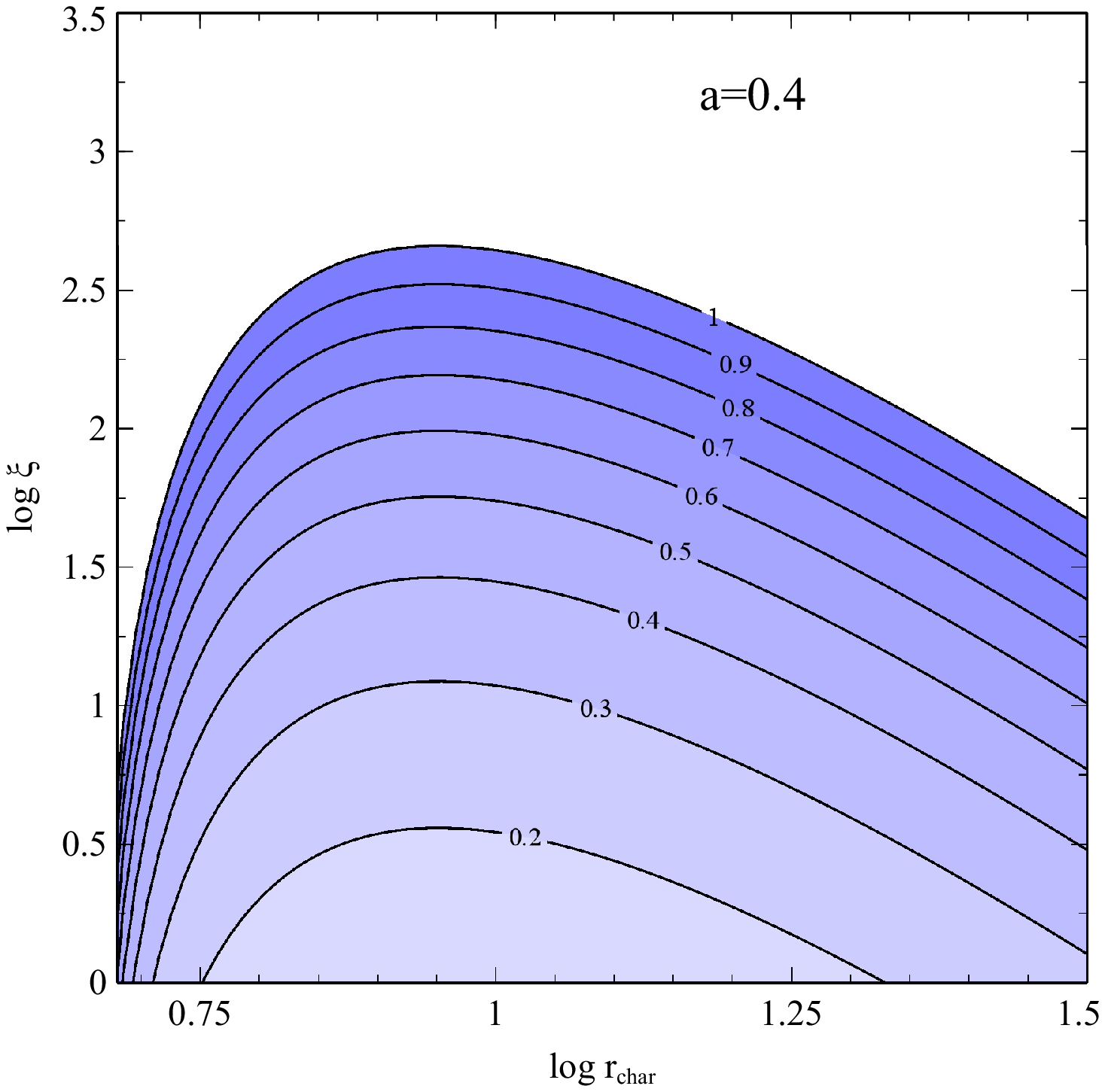,width=0.33\textwidth}
\psfig{figure=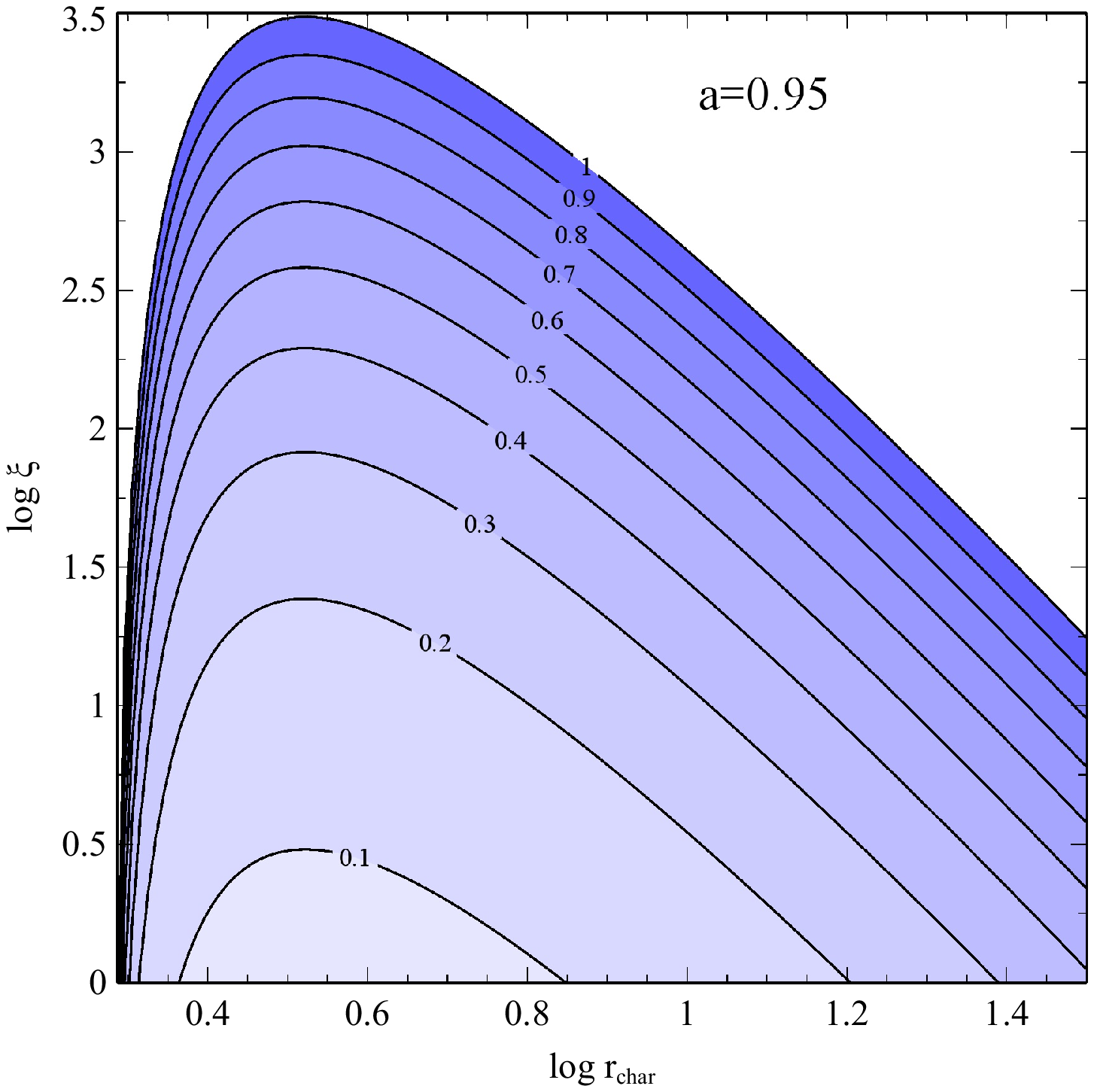,width=0.33\textwidth}
}
\caption{Contours of Eddington ratio $L_{\rm bol}/L_{\rm Edd}$ as functions of ionization parameter $\xi$ and characteristic reflection radius $r_{\rm char}$) for three different black hole spins, based on eqn~(2) of \cite{ballantyne:11a}.  This assumes a disk viscosity parameter of $\alpha=0.1$, an X-ray corona that radiates $f=0.1$ of the locally dissipated energy (although the result is very insensitive for $f=0.1-0.5$), and a radiative-efficiency corresponding to the binding energy at the ISCO.}
\label{fig:diskion}
\end{figure*}

The low metallicity found in \h18 is unusual compared with other well-studied AGN.  For example, it is lower by a factor of two than any of the (local) AGN studied in the \suzaku\ sample studies of \cite{walton:12a} or \cite{patrick:12a}.  This places constraints on the amount of star formation in the reservoir fueling this AGN that may be difficult to satisfy if the quasar is fueled by atomic or molecular gas.  We note that \cite{aravena:11a} has found significant amounts ($10^{10}\Msun$) of off-nuclear molecular gas in the core of this cluster, but no molecular gas has been detected in the nucleus \citep{combes:11a}.  The observed AGN metallicity is consistent, however, with accretion from the ICM itself as posited in the Compton-cooled ICM fueling scenario \citep{russell:10a,walker:14a}.   \cite{russell:10a} show that the core regions of the ICM have an iron abundance of $Z\sim 0.4Z_\odot$, the same as in the circumnuclear gas.  Furthermore, the inflowing ICM will be locked into the Compton-temperature of the quasar within about 5\,kpc \citep{russell:10a} provided that the density profile is shallower than $\rho(R)\propto R^{-2}$, preventing any fragmentation and star formation.  Of course, in order to produce the observed X-ray reflection signatures, some of this gas must cool within the immediate vicinity of the AGN.  This likely occurs due to self-shielding from the AGN radiation in a Compton-thick circumnuclear torus or outer accretion disk.

Under the assumption that the soft-excess is solely due to ionized disk reflection, our modeling sets some interesting constraints on fundamental parameters of this AGN.  If we naively identify $r_{\rm in}$ with the innermost stable circular orbit (ISCO), our upper limit of $r_{\rm in}<3.6r_g$ implies a lower limit on the black hole spin of $a>0.65$.  In fact, for the Eddington ratios likely applicable to \h18 (see below), the finite-thickness of the accretion disk may allow the reflecting region of the disk to penetrate somewhat into the ISCO \citep{reynolds:97a}; assuming a disk-thickness of $h\sim r_g$ softens our spin constraint to $a\gtrsim 0.4$ \citep[see Fig.~5 of ][]{reynolds:08a}.

Although model dependent, we can also estimate the Eddington fraction ($L_{\rm bol}/L_{\rm Edd}$) of \h18 from ionization state characterizing disk reflection.  This issue has been examined by \cite{ballantyne:11a} who employed the standard radiation-dominated \cite{novikov:73a} disk model to connect the disk ionization parameter through the disk density to the Eddington fraction.   In Fig.~\ref{fig:diskion} we use eqn.~(2) of \cite{ballantyne:11a} to calculate the Eddington fraction as a function of disk ionization and the characteristic radius of the disk reflection $r_{\rm chat}$ (i.e. the radius dominating the observed signal).  To apply this to \h18 (with $\log\xi=1.7$), we take $r_{\rm char}$ to be bracketed by the ISCO and our upper limit on the height of the illuminating source ($<11r_g$).  We see from Fig.~\ref{fig:diskion} that, unless $r_{\rm char}$ is very close to the ISCO, the measure ionization implies $L_{\rm bol}/L_{\rm Edd}\sim 0.5$.  We note \cite{ballantyne:11a} assume that the density of the disk atmosphere is the same as the mid-plane.  Any tapering of density into the disk atmosphere will lead to a decrease of the inferred Eddington fraction for a given pair of observables $(\xi,r_{\rm char})$.   A full calibration of this will require global radiation-pressure dominated magnetohydrodynamic (MHD) accretion disk simulations, but photospheric densities are plausibly an order of magnitude lower than mid-plane densities \citep{reynolds:08a} resulting in a revised constraint on the Eddington fraction of $L_{\rm bol}/L_{\rm Edd}\sim 0.25$.  With $L_{\rm bol}\approx 2\times 10^{47}\ergps$ \citep{russell:10a}, our estimated Eddington fraction of $L_{\rm bol}/L_{\rm Edd}\approx 0.25-0.5$ implies a black hole mass of $M_{BH}\approx 3-6\times 10^{9}\Msun$.   Using the parameters from \cite{russell:10a}, the Bondi accretion rate corresponding to this mass is $10-30\Msunpyr$, close to that needed to power this AGN.  

\section{Conclusions} 
\label{sec:conclusions}

{\it Suzaku} and {\it Chandra} have provided a unique glimpse into the workings of this rare BCG-hosted quasar, allowing us to isolate the spectral signatures of the AGN from those of the ICM.  X-ray reflection signatures reveal a sub-solar circumnuclear environment for this AGN, consistent with a picture whereby this quasar is accreting the ICM via a Compton-cooling driven cooling flow.  We also find a soft excess that can be described phenomenologically (with an ad-hoc black-body component) or as ionized reflection from the inner accretion disk.  Taking the accretion disk solution suggest a lower-limit to the black hole spin of $a>0.4$ and an Eddington fraction of $L_{\rm bol}/L_{\rm Edd}\approx 0.25-0.5$.

If quasar Compton-cooling is really responsible for the depressed entropy profile of the ICM core \citep{walker:14a}, the quasar would have had to have undergone an past outburst with $L_{\rm bol}\approx 4\times 10^{48}\ergps$.  To respect the Eddington limit during this outburst, \cite{walker:14a} argue that \h18 may possess an ultra-massive black hole, with $M_{BH}\sim 3\times 10^{10}\Msun$.   Our result suggests a slightly less massive black hole ($M_{BH}\approx 3-6\times 10^{9}\Msun$), necessitating the outburst of \h18 to have been super-Eddington by a factor of $\sim 5-10$.  Our mass estimate is significantly higher than that of \cite{dasyra:11a}, $M\approx 1\times 10^9\Msun$, based upon infrared emission line correlations.   It is maybe not surprising that \h18 does not follow normal emission line correlations given its unusual character.

\acknowledgments
\section*{Acknowledgments}
We thank Eric Miller for invaluable advice and discussions about XIS contamination and calibration.  CSR acknowledges support from the NASA-ADAP (grant NNX14AF86G), and AB from NSERC Discovery Grant Program.  

\bibliographystyle{jwapjbib}
\bibliography{chris}

\end{document}